\documentclass[a4paper,12pt]{article}
\usepackage[utf8]{inputenc}
\usepackage{overpic}
\usepackage{rotating}
\usepackage{algorithm}
\usepackage{algorithmic}
\usepackage{amsmath}
\usepackage{amssymb}
\usepackage{fancyhdr,lastpage}
\usepackage{enumerate}
\usepackage{graphicx}
\usepackage{epstopdf}
\usepackage[round]{natbib}
\usepackage{rotating}
\usepackage{mathrsfs}
\usepackage{tipa}
\usepackage{hyperref}
\hypersetup{colorlinks,allcolors=black}
\usepackage{authblk}

\newcommand{\bs}[1]{\boldsymbol{#1}}
\newcommand{\pd}[2]{\frac{\partial #1}{\partial #2}}

\newcommand{\ands}{ \ \ \ \mbox{ and } \ \ \ }
\newcommand{\spword}[1]{ \ \ \ \mbox{ {#1} } \ \ \ }

\begin{document}
\title{\vspace{-2cm}Implications of shallow-shell models for topographic relaxation on icy satellites}
\author[1]{\small Colin R. Meyer\thanks{colin.r.meyer@dartmouth.edu}}
\author[2,3]{Aaron G. Stubblefield}
\author[4]{Brent M. Minchew}
\author[5]{Samuel S. Pegler}
\author[6]{Alexander Berne}
\author[1]{Jacob J. Buffo}
\author[1]{Tara C. Tomlinson}
\affil[1]{Thayer School of Engineering, Dartmouth College, Hanover, NH 03755 USA}
\affil[2]{Earth System Science Interdisciplinary Center, University of Maryland, College Park, MD 20740 USA}
\affil[3]{Global Modeling and Assimilation Office, NASA Goddard Space Flight Center, Greenbelt, MD 20771 USA}
\affil[4]{Department of Earth, Atmospheric and Planetary Sciences, Massachusetts Institute of Technology, Cambridge, MA 02142, USA}
\affil[5]{School of Mathematics, University of Leeds, LS2 9JT, UK}
\affil[6]{Seismological Laboratory, California Institute of Technology, Pasadena, CA 91125 USA}
\maketitle

\begin{abstract}
Icy satellites host topography at many length scales, from rifts and craters on the small end to equatorial-pole shell thickness differences that are comparable to these bodies' circumference. The rate of topographic evolution depends on the rheology of the ice shell, the difference in density between the shell and potential global subsurface oceans, the ice shell thickness distribution, as well as the rate of tidal heat dissipation in the shell. The current paradigm is that icy satellites should not host stable small-scale topography. This idea comes from previous work using a ``shallow''-shell model (i.e. ice shell circumference is much larger than shell thickness) with a rigid outer crust. In this limit, large-scale topography relaxes over a longer time scale than small-scale features. Here we revisit this paradigm and analyze relaxation of topography starting from the Stokes equations for viscous fluid flow. We write out the shallow-shell models (with and without a rigid crust) and compare the resulting time scales to previous work. Using three-dimensional linearized flow for small-amplitude topographic perturbations, we recover the time scales of relaxation from the shallow models as well as shells that are not shallow. For a shell of constant viscosity, we find that the topographic relaxation time scale is constant for wavelengths larger than the ice thickness, an idea that runs contrary to the existing paradigm. For a shell with a viscosity that decreases exponentially with depth, we show numerically that there is a regime where the larger viscosity outer crust acts as a nearly rigid boundary. In this case, the relaxation time scale depends on the wavelength. For the largest spatial scales, however, the time scale becomes independent of wavelength again and the value is set by the average shell viscosity. However, the spatial scale that this transition occurs at becomes larger as the viscosity contrast increases, limiting the applicability of the scale-independent relaxation rate. These results for the relaxation of topography have implications for interpreting relaxed crater profiles, inferences of ice shell thickness from topography, and upcoming observations from missions to the outer solar system.
\end{abstract}

\section{Introduction}
Icy satellites are water ice-covered moons of the outer solar system. These icy moons occur around all of the gas giants, e.g. Jupiter hosts Europa, Saturn hosts Enceladus, Uranus hosts Miranda, Neptune hosts Triton, and many others. Ocean worlds are a subset of the icy satellites for which the ice shell overlies a liquid ocean. Europa is thought to have a liquid ocean due to magnetic reversal observed by the Galileo spacecraft \citep{Kiv2000}. Evidence for an ocean on Enceladus comes primarily from observations of the moon's libration \citep{Tho2016}. Some icy satellites are geologically active bodies hosting, for example, systems that suggest tectonics, plate motion, geysers, and convection \citep{Nim2016}. 

Large-scale topography, as is present on Enceladus \citep{Gie2008,Ies2014,Hem2018,Hem2019,Ber2023a,Ber2023b}, is expected to drive lateral differences in the melting temperature at the base of the ice shell \citep{Rem2022,Kan2022b,Sty2023}. These differences correspondingly imply differences in hydrostatic pressure, which may drive ocean circulation \citep{Ash2018,Zhe2021,Kan2022a,Kan2022b,Kan2022c}. The ice shell thickness also affects conductive and convective heat flow in the satellite \citep{Zhe2021,Car2021}. Knowledge of the ice shell thickness also informs the extent that surface is topography is isostatically compensated, e.g. Airy isostasy \citep{Hem2018,Hem2019, Cad2019a, Cad2019b, Aki2022}. In this way, improving dynamic ice flow models could improve inferences of ice shell thickness from observations of ice satellite's shape and gravity field. 

Given the central role large-scale ice flow plays in numerous processes and the upcoming gravity and topography data from upcoming planetary missions (e.g. NASA's Europa Clipper and ESA's Juice), understanding viscous deformation of icy satellite shells is important. In an earlier paper, \citet{Nim2004a} showed that there should not be any small-scale topography on icy satellites. In the next section, we write out the details of this argument mathematically, but in words, the \citet{Nim2004a} idea follows from a scaling analysis resulting in a relaxation time scale that is small for small-scale topography and large for large-scale topography. The implication is that small-scale topography relaxes away so quickly that only the large-scale topography is likely to be observable in the present. 

In this paper, we revisit the \citet{Nim2004a} argument for the purported non-existance of small-scale topography, make connections to viscous flow on glaciers and ice shelves on Earth, and present new scaling relationships for topographic relaxation on icy satellites. In section 2, we describe the model, boundary conditions, and two ``shallow'' approximations based on the fact that the ice shell circumference is much larger than its thickness, which is true for most icy satellites. In section 3, we present scalings for the different regimes and discuss the implications for topographic relaxation time scales. We also discuss a linearized analysis for perturbations to ice thickness. In section 4, we include spatial variation in the shell viscosity into our model and compare our results to the prior analysis. In section 5, we discuss the role of spatially varying viscosity in the context of prior work and our analysis. Finally, in section 6, we conclude by describing our results and comparing them to the scaling relationship derived by \citet{Nim2004a}.

\section{Model for ice flow within icy satellite shells}
We treat the flow within the ice shell as viscous motion at zero Reynolds number, where differences in ice shell thickness lead to gravity-driven deformation. The equations of conservation of momentum and mass in this context are the Stokes equations, given by
\begin{eqnarray}
\nabla \cdot \left\{\mu \left[ \nabla \bs{u}+\left(\nabla \bs{u}\right)^{T}\right] \right\} = \bs{\nabla}p +\rho g \bs{\hat{z}},\label{eqn:Stokes1}\\
\bs{\nabla}\cdot\bs{u}=0,\label{eqn:Stokes2}
\end{eqnarray}
where $\bs{u}$ is the velocity of ice parcels within the shell, $p$ is the pressure, $\rho_i$ is the ice density, which we treat as constant, $g$ is gravity, and $\mu$ is the ice viscosity. The Stokes equations for a Newtonian fluid are time-invariant, reversible, and linear \citep{Ach1990}. We consider nonlinear effects in section \ref{sec:tempdep}. Laboratory, field, and remote sensing studies of glacier ice \citep[e.g.][]{Gol2001,Mil2022} show that the ice viscosity is a nonlinear function of the strain rate (i.e., a power law, which is sometimes called Glen's law) and a strongly nonlinear function of the temperature (Arrhenius exponential). There has been significant work on exploring the roles of nonlinear processes on the viscosity, e.g., temperature, grain size, and strain rate \citep{Nim2004a,Bar2005,Rud2012,Cad2019b,Cad2019a}. Here we focus on a constant ice viscosity in the first part of the paper and a depth-dependent viscosity in the second part of the paper. 


\begin{figure}
\centering
\includegraphics[width=\linewidth]{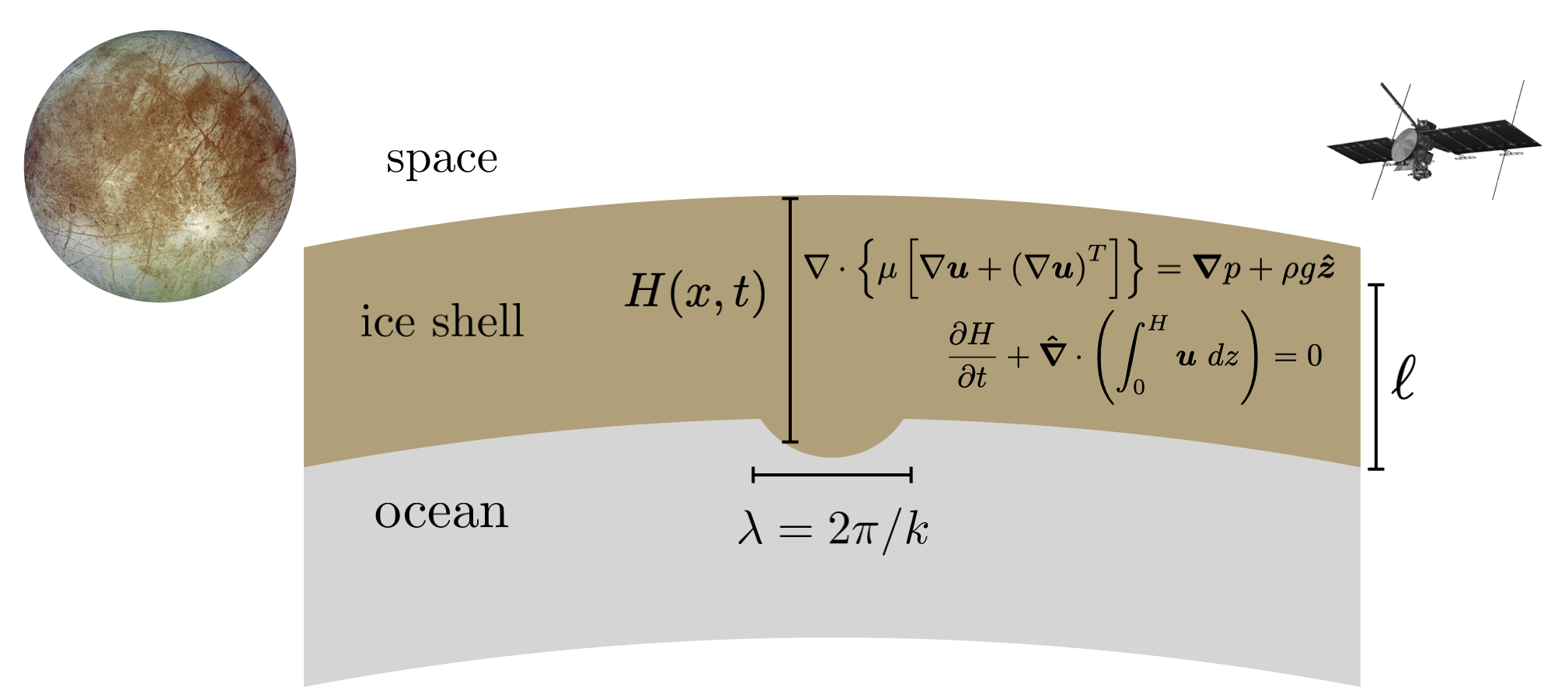}
\caption{Schematic of topography on an icy satellite. Equations shown for the evolution of Newtonian viscous ice with shell thickness $H$. The parameters and equations shown on the figure are described in the text.}
\label{fig:schematic}
\end{figure}

For Stokes flow applied to the ice shell, the boundary conditions at the base ($z=b$) are buoyancy in the vertical and stress free in the horizontal, i.e. 
\begin{equation}
\bs{\sigma}\cdot\bs{n} = \rho_w g (\ell - s) \bs{n}\spword{at}z=b,
\end{equation}
where $\bs{\sigma}$ is the stress tensor, $\bs{n}$ is the normal vector to the interface, $\rho_w$ is the ocean density, and $\ell$ is the height of flotation \citep[see schematic in figure \ref{fig:schematic};][]{Stu2023a}. We take $\ell = \rho_i H_0/\rho_w$, to indicate the level that the water would rise to if there was an opening from top to bottom. Here $H_0$ is the initial ice shell thickness. At the surface $z=s$, the boundary condition is a stress-free condition, given by
\begin{equation}
\bs{\sigma}\cdot\bs{n} = 0\spword{at}z=s,
\end{equation}
but some authors, including \citet{Nim2004a}, use a no slip condition, of the form
\begin{equation}
\bs{u} = 0\spword{at}z=s,
\end{equation}
because the surface is cold and likely rigid. By imposing this condition, the surface cannot move laterally or vertically. We explore the difference between these two boundary conditions in the next section. 

Integrating mass conservation, equation \eqref{eqn:Stokes2}, from $z=b$ (base) to $z=s$ (surface), gives 
\begin{equation}
\int_{b}^{s}{\bs{\nabla}\cdot\bs{u}}~dz = w(s)-w(b) +  \int_{b}^{s}{\bs{\hat{\nabla}} \cdot\bs{\hat{u}}}~dz = 0,\label{eqn:intmasscon}
\end{equation}
where the hat diacritic indicates that the vector components are only in the horizontal direction, i.e. $\bs{\hat{u}} = (u,v)$ and $\bs{\hat{\nabla}} = (\partial_x,\partial_y)$. In the case that there are stress-free conditions at the top and bottom of the shell, we have a kinematic condition that relates the vertical velocities to the interface motion, as in
\begin{equation}
w(s) = \pd{s}{t}+\bs{\hat{u}}\cdot\bs{\hat{\nabla}}s\ands w(b) = \pd{b}{t}+\bs{\hat{u}}\cdot\bs{\hat{\nabla}}b.
\end{equation}
In the no slip top boundary case, $w(s) = 0$. We have neglected the effects of melting or freezing and impact cratering on the basis that they are small perturbations that set up topographic variations but are not important for the viscous relaxation. Inserting these kinematic conditions into equation \eqref{eqn:intmasscon}, writing $H = s-b$, and using the Leibniz integral rule results in 
\begin{equation}
\pd{H}{t}+\bs{\hat{\nabla}}\cdot \int_{b}^{s}{\bs{\hat{u}}}~dz  = 0,\label{eqn:depthintegralmc}
\end{equation}
which is an evolution equation for the shell thickness $H$. This is the form of mass conservation that the shallow models we describe in the next section use as a starting point.

\subsection{Shallow models for ice shell flow}
To analyze topographic relaxation, \citet{Nim2004a} starts from a reduced form of equations \eqref{eqn:Stokes1} and \eqref{eqn:Stokes2}. The limit Nimmo that considers arises because the ice shell is ``shallow'', meaning that the shell depth is much smaller than the circumference, and because the surface is taken to be a ``rigid lid'', meaning that $\bs{u}=0$ at the surface, i.e. there is no slip. In glaciology, this asymptotic reduction of the Stokes equations is referred to as the ``shallow ice approximation'' or SIA \citep[see derivation in][]{Hew2021}. The shallow ice approximation applies to glacier flow where the ice is frozen to the ground below and the dominant force balance is between the driving stress (gradient in ice thickness multiplied by density and gravity) and shearing viscous deformation of the ice. In this limit, the pressure distribution within the ice is nearly hydrostatic, causing variation in topography to lead to lateral pressure gradients that drive flow, where ice flows from thick to thin regions. In the context of the ice shell, the shallow ice approximation implies that there is no flow at the surface and the shear within the ice shell balances the differences in ice shell thickness. Using the variables shown schematically in figure \ref{fig:schematic}, we can rewrite equation \eqref{eqn:depthintegralmc} for the case of the shallow ice approximation as
\begin{equation}
\pd{H}{t} = \bs{\hat{\nabla}}\cdot\left( \frac{\rho_i\Delta \rho g H^3}{3\mu \rho_w}\bs{\hat{\nabla}}H\right),
\label{eqn:sia}
\end{equation}
where $H$ is the evolving ice shell thickness, $\Delta \rho = \rho_w-\rho_i$ is the density difference, $g$ is gravity, and $\mu$ is the ice viscosity, which we treat as Newtonian for illustration. In this way, the shallow-ice approximation is a nonlinear diffusion equation for the evolution of the ice shell thickness with time \citep{Bue2005}. \citet{Nim2004a} uses a non-Newtonian ice rheology \citep[e.g.,][]{Gle1955,Gol2001} but then writes out the time scale for a Newtonian viscosity. In addition to \citet{Nim2004a}, the shallow ice approximation has also been applied to icy satellites by, for example, \citet{Ash2018} as well as \citet{Kan2020}.

The shallow ice approximation, however, is not the only shallow reduction of the Stokes equations that could be applied to the shells of icy satellites. Given that there are no stress constraints on the surface of the ice shell, a stress-free boundary condition is likely more appropriate than a no-slip condition, regardless of how viscous the cold surface ice becomes. With stress-free conditions at the top and bottom of the shell, the driving stress can no longer be balanced by shear within the shell and now must be matched with the extensional stress. This dynamic regime occurs in Earth's ice shelves and is known as the ``shallow shelf approximation'' or SSA in glaciology \citep{Mor1987,Mac1989}. For a Newtonian fluid, a derivation of SSA is given in \citet{Peg2012} and the equations are
\begin{eqnarray}
\pd{H}{t}+\bs{\hat{\nabla}}\cdot\left( H\bs{\hat{u}}\right) &=& 0,\label{eqn:ssamc}\\
\bs{\hat{\nabla}}\left( H \bs{\hat{\nabla}} \cdot \bs{\hat{u}}\right) + \bs{\hat{\nabla}}\cdot \left( H \dot{\bs{\epsilon}}\right) &=& \frac{\rho_i g \Delta \rho }{2\mu \rho_w} H\bs{\hat{\nabla}}H,\label{eqn:ssafb}
\end{eqnarray}
where $H$ is the ice shell thickness, $\bs{\hat{u}}$ is the horizontal velocity vector and $\dot{\bs{\epsilon}} = [\bs{\hat{\nabla}}\bs{\hat{u}}+\bs{\hat{\nabla}}\bs{\hat{u}}^{T}]/2$ is the horizontal strain rate tensor.


\section{Results for the time scales of topographic relaxation} 
\subsection{Scaling analysis for the relaxation time}
\begin{figure}
\centering
\includegraphics[width=0.6\linewidth]{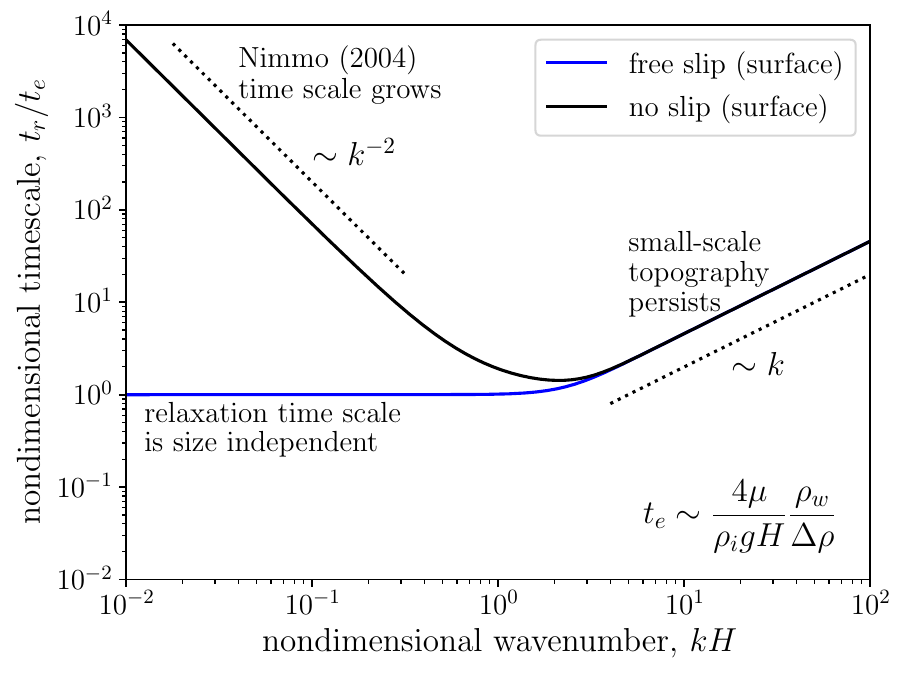}
\caption{Linearized Stokes flow calculations from \citet{Stu2023b,Stu2023a} demonstrating the scalings and regimes for basal topography. In the no slip (rigid surface boundary condition) case, shallow-ice prevails and there is an increase in the relaxation time for large-scale topography \citep[equation \eqref{eqn:siascaling}; ][]{Nim2004a}. For free slip (stress-free, viscous relaxation at the surface), the shallow-shelf approximation indicates that large-scale topography relaxes at the same rate as small-scale topography, i.e. equation \eqref{eqn:tefromscaling}. For disturbances less than the shell thickness, both models show that the relaxation time increases, as in equation \eqref{eqn:smallscales}.}
\label{fig:ags_timescale}
\end{figure}

In order to build intuition, we start by scaling the shallow-ice and shallow-shelf approximations. We consider topography of a given scale with a wavenumber $k = 2\pi/\lambda$, where $\lambda$ is the wavelength. In either of the shallow limits, the topography is broader than the ice thickness, so that $kH\ll 1$, so our scalings give the small-wavenumber (long-wavelength) relaxation time scales. For shallow-ice flow, topography will relax according to equation \eqref{eqn:sia}. Analyzing the dimensions of the terms in \eqref{eqn:sia}, we see that 
\begin{equation}
\frac{H}{t_s} \sim \frac{\rho_i\Delta \rho g H^4}{3\mu \rho_w}k^2,
\end{equation}
where $t_s$ is the time scale for relaxation, i.e.
\begin{equation}t_s \sim \frac{\mu}{\Delta \rho g k^2 H^3},\label{eqn:siascaling}\end{equation}
as described in \citet{Nim2004a} for a no-slip surface. The inverse square dependence of relaxation time on wavenumber means that topography on the largest scale will relax over a longer time than the topography that is at a smaller scale (but still larger than the ice thickness). We show this scaling relationship in figure \ref{fig:ags_timescale}. 

For shallow-shelf flow, where there is a stress-free surface, topography will relax according to equations \eqref{eqn:ssamc} and \eqref{eqn:ssafb}. Again considering topography with a wavenumber $k$, we scale these equations to find 
\begin{eqnarray}
\frac{H}{t_s} &\sim& k H u ,\label{eqn:ssamc_scale}\\
k^2 H u &\sim& \frac{\rho_i g \Delta \rho }{2\mu \rho_w} k H^2.\label{eqn:ssafb_scale}
\end{eqnarray}
Rearranging we find that the timescale for relaxation is 
\begin{equation}\hspace{0.08cm}t_s \sim \frac{2\mu \rho_w}{\rho_i \Delta \rho g H },\label{eqn:tefromscaling}\end{equation}
which is independent of the wavenumber.

For disturbances that are smaller than the shell thickness, i.e. $kH\gg1$ or $\lambda\ll H$, then we consider the Stokes equations with stress-free conditions on the top and bottom, i.e. equations \eqref{eqn:Stokes1}-\eqref{eqn:depthintegralmc}. We consider a topographic disturbance at the bottom of the shell. Taking $\bs{u}=(u,v,w)$, we have no shear stress at the base $z=s$, i.e.
\[\sigma_{xz} = \pd{u}{z}+\pd{w}{x} = 0,~~~\sigma_{yz} = \pd{v}{z}+\pd{w}{y} = 0 \ands \bs{\nabla}\cdot\bs{u}=\pd{u}{x}+\pd{v}{y}+\pd{w}{z}=0.\]
From scaling, the only way to make these compatible is for $u\sim v\sim w$ and $x\sim y\sim z \sim 1/k$. The normal stress and kinematic condition at this interface give that 
\[\sigma_{zz} = -p + 2\mu \pd{w}{z} = -\Delta \rho g s\ands \pd{s}{t} = w,\]
These gives the scalings 
\[2 \mu k w \sim \Delta \rho g s \ands t_s \sim \frac{s}{w},\]
from which we find
\begin{equation}t_s \sim \frac{2\mu}{\Delta \rho g H} kH,\label{eqn:smallscales}\end{equation}
which increases linearly with wavenumber $k$ \citep{Tur2002} and is independent of the shell thickness $H$. As shown in figure \ref{fig:ags_timescale}, this scaling relationship applies to both no-slip and free-slip surface boundary conditions. 

\subsection{Linearized equations for topographic perturbations\label{sec:linear}}
We now describe the linearized model for topographic perturbations to a floating viscous shell, as introduced by \citet{Stu2023b,Stu2023a}. Starting from a parallel-sided ice shell of thickness $H$ with $b = 0$ and $s=H$, we can write out a base case of (i) zero flow, and (ii) hydrostatic pressure, which satisfies the Stokes equations \eqref{eqn:Stokes1} and \eqref{eqn:Stokes2} subject to boundary conditions. We then look at the perturbed Stokes equations, where each variable $\phi$ is given as the value in the base case $\overline{\phi}$ plus a small perturbation $\phi_*$, where $\phi_*\ll \overline{\phi}$. Inserting the variables into the Stokes equations and keeping only the terms that are linear in the perturbed variables, we have that 
\begin{eqnarray}
\mu \bs{\nabla}^2\bs{u_*} &=& \bs{\nabla}p_*,\\
\bs{\nabla}\cdot\bs{u_*}&=&0,
\end{eqnarray}
where gravity drops out because of the hydrostatic base case. The viscosity is constant in the base case, but also results in a constant in the perturbed case \citep{Stu2023b}. Linearizing the stress boundary conditions from before, we have that 
\begin{eqnarray}
\pd{s_*}{t} = w,~~~p_* - 2\mu \pd{w_*}{z} = \rho_i g s_*,\spword{at}z=s,\label{eqn:bc1}\\
\pd{b_*}{t} = w,~~~p_* - 2\mu \pd{w_*}{z} = -\Delta \rho g b_*,\spword{at}z=b,\label{eqn:bc2}\\
\pd{u_*}{z}+\pd{w_*}{x} = 0,~~~\pd{v_*}{z}+\pd{w_*}{y} = 0,\spword{at}z=b,~s.\label{eqn:bc3}
\end{eqnarray}

Since we have linearized the equations, we can look for a Fourier transform solution for arbitrary topographic perturbations of wavenumber $k$. The details of the derivation are shown in \citet{Stu2023b,Stu2023a}, but the result is two coupled ordinary differential equations for the evolution of the surface and base perturbations, i.e.
\begin{eqnarray}
t_\textrm{r} \pd{\widehat{s}}{t} + \mathsf{R} \widehat{s} &=& -\mathsf{B}\delta \widehat{b},\label{eqn:ode1}\\
t_\textrm{r} \pd{\widehat{b}}{t} + \delta \mathsf{R} \widehat{b} &=& -\mathsf{B} \widehat{s},\label{eqn:ode2}
\end{eqnarray}
where $\delta = \Delta \rho / \rho_i$ and the time scale $t_\textrm{r}$ is given by
\begin{equation}
t_\textrm{r} = \frac{2\mu}{\rho_i g H}.
\end{equation}
The two transfer functions $\mathsf{R}$ and $\mathsf{B}$ are functions of the wavenumber $k$ and given by
\begin{eqnarray}
\mathsf{R} &=& \frac{e^{4{kH}} +4{kH} e^{2{kH}} -1 }{kH\{e^{4{kH}} -2[1+2(kH)^2]e^{2{kH}} +1\}}\label{eqn:R},\\
\mathsf{B} &=&\frac{ 2({kH}+1)e^{3{kH}}+2({kH}-1)e^{{kH}} }{kH\{e^{4{kH}} -2[1+2(kH)^2]e^{2{kH}} +1\}},\label{B}
\end{eqnarray}
where $\boldsymbol{k} = [k_x,k_y]^\mathrm{T}$ is the horizontal wavevector 
with magnitude $k=\sqrt{k_x^2+k_y^2}$.

Writing the system of ordinary differential equations \eqref{eqn:ode1} and \eqref{eqn:ode2}, as a matrix equation, we can find the eigenvalues $\lambda_{\pm}$, which are the time scales for evolution. The eigenvalues for this system are given as
\begin{eqnarray}
\lambda_\pm = \frac{\delta+1}{2t_\textrm{r}} \mathsf{R} \pm \frac{\sqrt{4\delta\mathsf{B}^2 +  \mathsf{R}^2(\delta-1)^2}}{2t_\textrm{r}},\label{eqn:lambda}
\end{eqnarray}
and the time scale for topographic relaxation are given as 
\begin{eqnarray}
t_{\pm} = \frac{1}{\lambda_{\pm}},
\end{eqnarray}
where the plus sign corresponds to the relaxation of basal disturbances and the minus sign is the relaxation of surface disturbances \citep{Stu2023a}. Here we focus on the basal case because the surface relaxation follows the basal timescale for large-scale deformation, as described in the next section.

\begin{figure}[!ht]
\centering
\includegraphics[width=\linewidth]{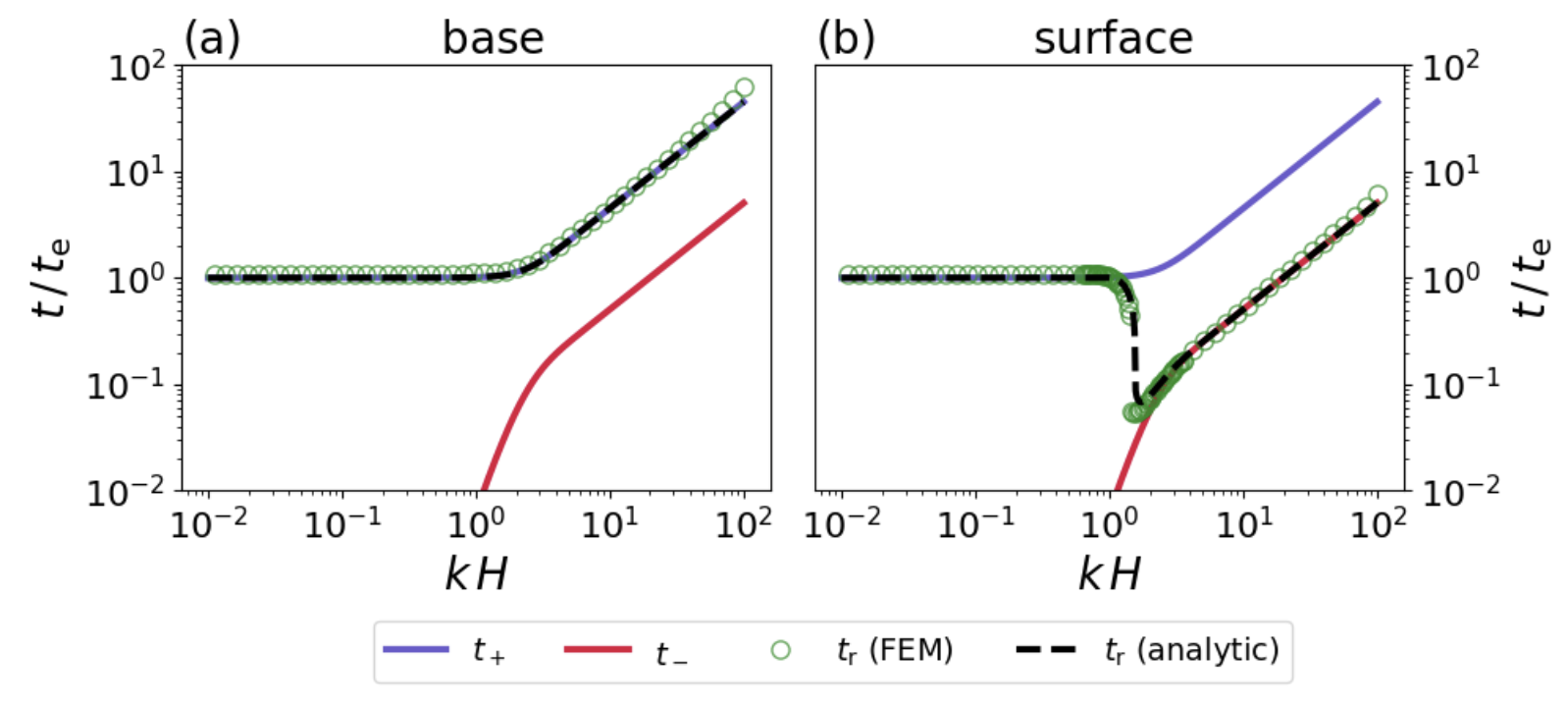}
\caption{Time scale for relaxation for a Newtonian ice shell with a free-slip (stress-free) surface boundary condition. Comparison between linearized theory from \citet{Stu2023b,Stu2023a} and our finite element simulations. (a) perturbations to the base, where the relaxation curve follows $t_+$ for all wavenumbers. The green dots and black dashed line are the same as the blue line in figure \ref{fig:ags_timescale}. (b) perturbations to the surface, where the relaxation curve follows $t_+$ for small wavenumbers then transitions to $t_-$ for large wavenumbers, due to hydrostatic adjustment.}
\label{fig:femtest}
\end{figure}

In the long wavelength (small wavenumber) limit as $k\rightarrow0$, \citet{Stu2023a} shows that the basal relaxation time scale is given by
\begin{equation}
t_+ \equiv t_e = 2t_\textrm{r} \left(1+\frac{1}{\delta}\right) = \frac{4\mu}{\Delta \rho g H}\frac{\rho_w}{\rho_i}.
\end{equation}
The time scale for relaxation of the surface perturbation $t_{-}$ becomes very small, following as scaling of $t_{-} \sim k^{-4}$. These results are shown in figure \ref{fig:femtest}. Importantly, the basal perturbations relax at a rate given by $t_e$ which is the same (modulo a factor of 2) as the time scale derived from scaling the shallow-shelf approximation, i.e. equation \eqref{eqn:tefromscaling}. 

\section{Viscosity variation within the shell\label{sec:tempdep}}
We now consider the role of the temperature-dependence of viscosity on the linear analysis we described in section \ref{sec:linear}. Considering small perturbations away from a hydrostatic shell at rest, the base case is independent of viscosity. We take the viscosity to only be a function of depth, which is a simplified choice for the case of a conductive shell \citep{Nim2004a}, given by
\begin{equation}
\mu(z) = \mu_0 \exp\left\{z/d\right\},
\end{equation} 
where $d$ is the scale over which the viscosity changes within the shell. Because the problem is nonlinear and derivatives of the viscosity result in new shear terms, we cannot develop a linearized analysis along the lines of \S\ref{sec:linear}. Therefore, we solve the equations \eqref{eqn:Stokes1} and \eqref{eqn:Stokes2} numerically using the finite element method developed by \citet{Stu2023b,Stu2023a}. To start we reproduce our analysis for the constant viscosity ice (figure \ref{fig:ags_timescale}). We compute the relaxation timescale of sinusoidal perturbations of a given wavelength. We consider sinusoidal surface and base perturbations separately. We determine the timescale by computing an exponential fit to the decay of maximum topography with time. 
In essence, this is the time for the perturbations to reach $1/e$ of their original height, which is consistent with the linearized analysis in \S\ref{sec:linear}. The result is shown in figure \ref{fig:femtest}, where we plot the theoretical decay rates and finite element simulation results. 

We see that the basal topographic disturbances follow the $t_+$ line for all wavenumbers whereas the surface topography relaxes along the $t_+$ line for small wavenumbers and then jumps to the $t_-$ line for large wavenumbers. The reason for the jump is that buoyancy allows surface topography that is out of hydrostatic equilibrium to relax faster than basal perturbations by restoring the hydrostatic balance. 

\begin{figure}[!ht]
\centering
\includegraphics[width=0.9\linewidth]{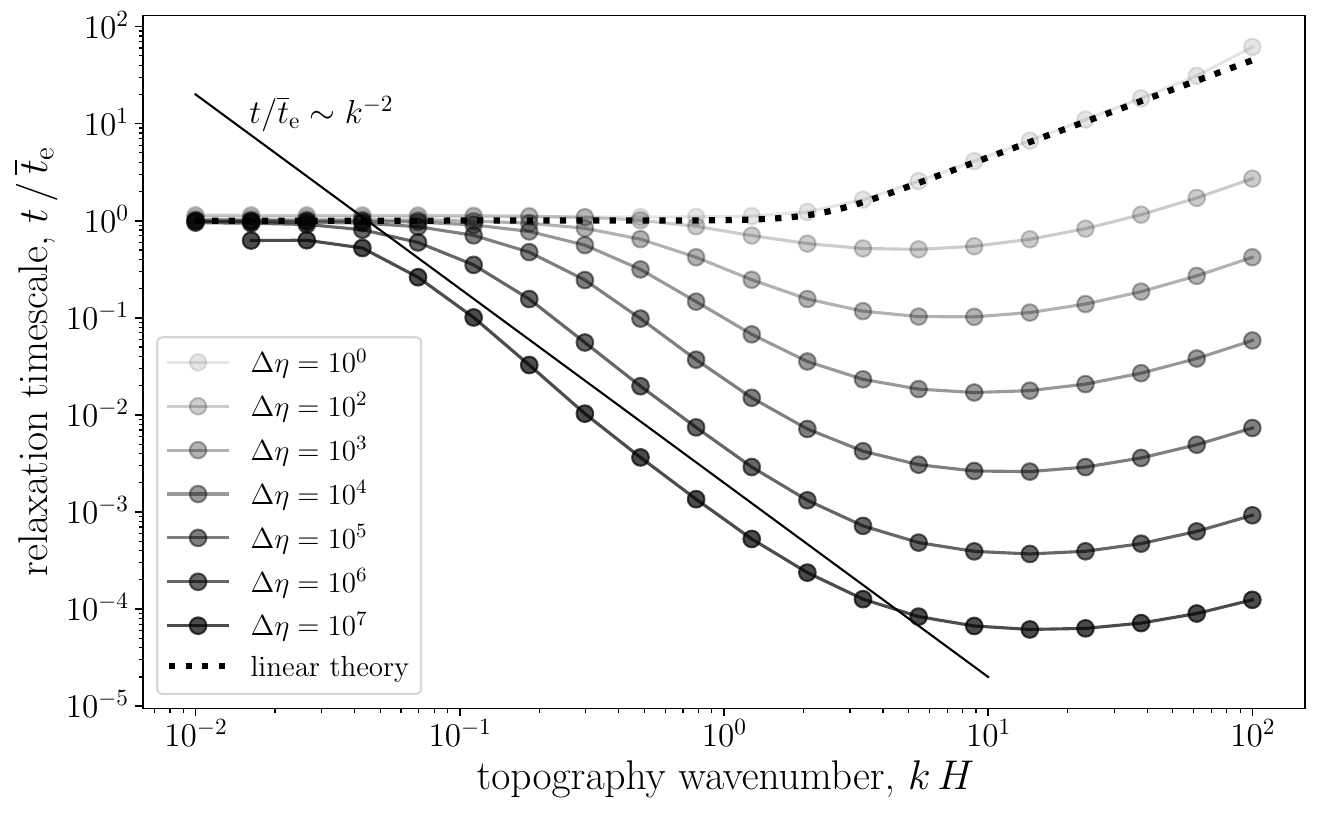}
\caption{Relaxation timescale for basal perturbations with variation in viscosity. The linear theory, i.e., the dotted line, follows from figures \ref{fig:ags_timescale} and \ref{fig:femtest} \citep{Stu2023b,Stu2023a}. The effect of increasing viscosity is to generate a regime of effectively shallow ice approximation flow ($t/\overline{t}_e \sim k^{-2}$) where there is a no-slip boundary condition at the surface. The transition between shallow ice flow (no slip) and shallow shelf (free slip) occurs at smaller and smaller wavenumbers as the viscosity contrast increases.}
\label{fig:stubavgvisc}
\end{figure}

Now with confidence that our finite-element calculations can reproduce the relaxation time scale in constant viscosity case, we use the same finite-element model to calculate the relaxation time scale with a depth-dependent viscosity. In figure \ref{fig:stubavgvisc}, we show the resulting relaxation timescales as a function of wavenumber for increasing viscosity ratio given by
\begin{equation}\Delta \mu = \mu(H)/\mu_0 = \exp\left\{H/d\right\}.\end{equation}
The timescale $t$ is normalized by the relaxation based on the average viscosity $\overline{\mu}$, i.e.
\begin{equation}
\overline{\mu} = \frac{1}{H}\int_{b}^{s}{\mu~dz} = \mu_0 \frac{\Delta \mu-1}{\ln(\Delta \mu)}.\label{eqn:mubar}
\end{equation}

The structure of the curves contains a few similarities and differences to the constant viscosity results. At large wavenumbers, the relaxation timescale for surface perturbations increases with wavenumber, and the magnitude increases as the viscosity increases. In other words, small-scale surface topography relaxes more slowly than the constant viscosity case. For basal perturbations at large wavenumbers, the opposite is true: the relaxation timescale increases with wavenumber but decreases with viscosity, meaning that small-scale basal topography relaxes more quickly than the constant viscosity case. At small wavenumbers, the timescales converge to $\overline{t}_e$. In essence, the timescale arguments for the shallow ice approximation articulated in the first part of the paper hold, only now for the average viscosity $\overline{\mu}$. So for topography at scales larger than the ice shell thickness, we can expect it to relax over a timescale of 
\begin{equation}
\overline{t}_e \sim \frac{4\overline{\mu} \rho_w}{\rho_i \Delta \rho g H},\label{eqn:avemu}
\end{equation}
which increases as $\overline{\mu}$ increases, but the dynamics stay the same: the shallow-shell approximation results in a wavenumber-independent relaxation time scale. 

However, there is an important middle regime that emerges in the simulations shown in figure \ref{fig:stubavgvisc}. For wavelengths around the ice thickness ($kH \sim 1$) and below, the relaxation time scale follows a wavenumber-dependent trend with a $t/\overline{t}_e \sim k^{-2}$ scaling, showing that it is relaxing according to the shallow-ice approximation (SIA). This middle, SIA regime is more pronounced for larger viscosity contrasts and extends to larger length scales (smaller wavenumbers). It appears that the stiffer, larger viscosity ice near the surface is acting as a rigid plug, reminiscent of simulations of icy satellite convection, and allows for effectively a no slip condition to prevail at the surface. 

\section{Discussion}
To see why the timescales converge on $\overline{t}_e$, we look back to the shallow-shelf approximation \citep{Mor1987,Mac1989}. For our analysis, we follow \citet{Peg2012}. Starting from their equation (2.18), where the depth-integrated force balance is written as
\begin{equation}
\bs{\hat{\nabla}} \cdot \left[ \int_{b}^{s} \bs{\sigma}\hspace{0.05cm}dz\right] = - \frac{\rho_i^2 g}{\rho_w} H\bs{\hat{\nabla}}H,\label{eqn:peg2.18}
\end{equation}
which we can rationalize as a continuum force balance, e.g. equation \eqref{eqn:Stokes1}, where hydrostatic pressure has been depth integrated. Now to proceed with the derivation we use the definition of the stress tensor, i.e.,
\begin{equation}
\bs{\sigma} = -p \bs{I} + 2\mu \bs{\dot{\epsilon}},\label{eqn:defofsigma}
\end{equation}
 and the pressure, which is given by
\begin{equation}
p = -\rho g (z-h) - 2\mu\left(\bs{\hat{\nabla}}\cdot\bs{\hat{u}}\right),\label{eqn:pressure}
\end{equation}
where the viscosity $\mu$ can be a function of the strain rate $\dot{\epsilon}_e$ and the temperature $T$. Inserting the pressure equation \eqref{eqn:pressure} into the stress definition equation \eqref{eqn:defofsigma}, then into the depth-integrated force balance \eqref{eqn:peg2.18}, and then integrating we find that
\begin{equation}
\bs{\hat{\nabla}}\left( \overline{\mu} H \bs{\hat{\nabla}} \cdot \boldsymbol{\hat{u}}\right) + \bs{\hat{\nabla}}\cdot \left( \overline{\mu} H \dot{\boldsymbol{\epsilon}}\right) = \frac{\rho_i g \Delta \rho }{2\rho_w} H\bs{\hat{\nabla}}H,\label{eqn:shallowshellrheology}
\end{equation}
where the depth-integrated viscosity $\overline{\mu}$ is given by equation \eqref{eqn:mubar}. Thus, the force balance for the shallow-shell model with a rheology that depends on strain rate and temperature reduces to the same shallow-shelf balance as before, i.e. equation \eqref{eqn:ssafb}, except the viscosity is now the depth-integrated average viscosity. This result shows that ice rheology only qualitatively modifies the timescale for topographic relaxation in the shallow-shell limit. In particular, adding a rheological dependence on strain rate and temperature does change the result that topographic relaxation is independent of the wavelength of the topography. Equations \eqref{eqn:shallowshellrheology} and \eqref{eqn:avemu} do not depend on the strength at which the rheology changes with temperature. In other words, even if the rheology contrast is six orders of magnitude, the leading order shallow Stokes flow dominated by extensional stress is still given by equation \eqref{eqn:shallowshellrheology}. This explains why the surface of an icy satellite could be cold and brittle, yet still participate in the viscous relaxation of extensional stresses.

\begin{figure}[!ht]
\centering
\includegraphics[width=0.8\linewidth]{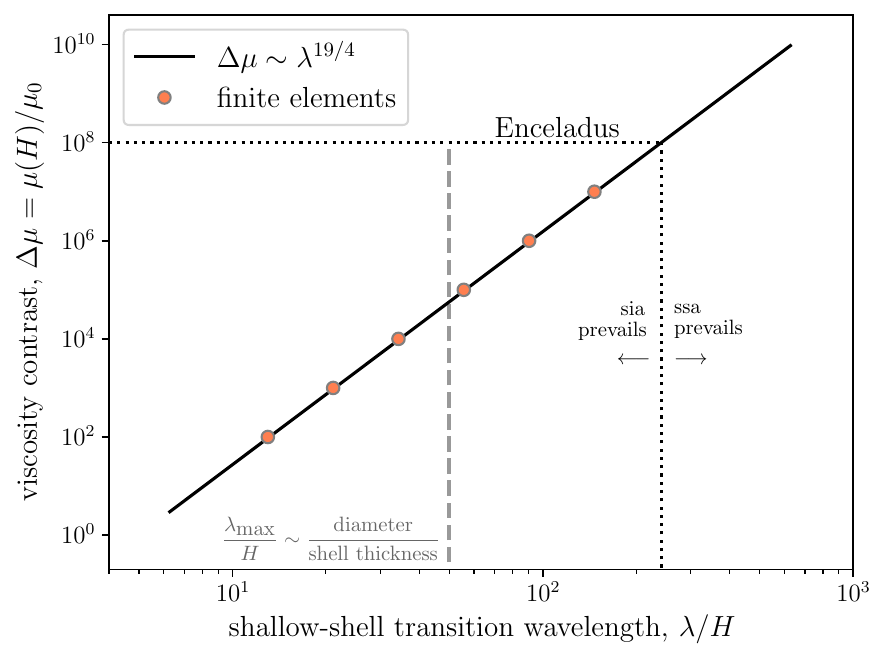}
\caption{A key feature of including a depth-dependent viscosity is that a regime that resembles the shallow ice approximation develops for intermediate wavenumbers. Here we show the topographic scale of the transition between SIA and SSA (i.e., where $t/\overline{t}_e\rightarrow 1$). As the viscosity contrast increases, the minimum size of the topographic variation that obeys SSA also increases. Disturbances smaller than this scale follow a version of SIA. The power law fit is empirical.}
\label{fig:transition_topo}
\end{figure}
However, as the viscosity contrast increases, the scale of topography that is independent of wavelength also increases (figure \ref{fig:stubavgvisc}). We quantify this effect by examining the wavenumber where the relaxation time scale approaches the shallow-shelf approximation 
(SSA) relaxation timescale, equation \eqref{eqn:avemu}. In other words, for wavenumbers where $t/\overline{t_e} \rightarrow 1$ and below, the topography will relax as a stress-free shallow shell. We plot the results for the transition wavelength in figure \ref{fig:transition_topo}. We find that the transition wavelength increases as a power-law with the viscosity contrast $\Delta \mu$, as given by
\begin{equation}
\Delta \mu \simeq 3\left(\frac{\lambda}{H}\right)^{19/4},
\end{equation}
where the power-law fit is empirical. Thus, for topographic evolution at scales larger than this wavelength, the deformation will follow SSA. For smaller scales (yet larger than the ice thickness), the shallow-ice approximation holds. 

As an example, Enceladus likely has a viscosity contrast of $\Delta \mu \sim 10^{8}$ Pa$\cdot$s \citep{Beh2015}. The largest scale motions would be close to the size of the moon, i.e. $\sim500$ km and the ice thickness is on the order of 10 km \citep{Par2024}. In this way, as shown schematically on the figure, the deformation of Enceladus' shell should follow the shallow-ice approximation, as described by \citet{Nim2004a}. Enceladus' exact viscosity contrast is less important here, since it would need to be on the order of $10^5$ Pa$\cdot$s for it to follow the shallow-shelf approximation. A similar analysis for Europa shows that the topography will also relax according to the shallow-ice approximation. In order for a moon to experience a regime of shallow-shelf deformation, the length scale of the topography must be large, the ice shell must be thin, and the viscosity contrast must be small. These requirements limit the number of moons in the solar system where SSA would be applicable. 

Sufficiently high-resolution shape models \citep[e.g.,][]{Sch2024,Par2024} allow for spectral analyses of topographic relaxation at icy satellites. Generally, the variances of topography per spherical harmonic degree of planetary bodies follow power-law distributions (i.e. Kaula’s rule) and deviations from this behavior (e.g., breaks) are believed to arise due to a transition from elastically supported topography (at short wavelengths) to compensated topography  \citep[at long wavelengths;][]{Ara2009}. Here we find that, in fact, two breaks in topographic spectra should occur for icy satellites with viscosity structures that vary with depth: one at long-wavelengths corresponding to the transition from behavior associated with the SIA to that associated with the SSA and a second, more subtle transition in which relaxation shifts from a dependence on $k^{-2}$ to $k$ (as in Figure \ref{fig:stubavgvisc}). At Enceladus and Europa, the presence of these spectral breaks across the crust rely on the presence of a very low viscosity subsurface layer and would correspondingly indicate the presence of global subsurface oceans. Fitting variance spectra to model predictions (e.g., Figure \ref{fig:stubavgvisc}) may also constrain the viscosity structure and thickness of the ice shells of these bodies in the future.

\section{Conclusions}
To understand topographic evolution on icy satellites we consider both constant and spatially varying viscosity. For the first part of the paper, we focused on a constant viscosity. We found that in the shallow-shelf approximation, basal features larger than the shell thickness relax at the same rate regardless of size. Topography at scales smaller than the ice shell thickness relax more slowly, regardless of a no-slip versus free-slip surface. For the second part of the paper, we analyzed the role of viscosity variation in the shell. We quantified the viscosity structure with an exponential increase and characterized the variation by the ratio of the viscosity at the surface to the viscosity at the base, then we performed finite-element simulations of topographic relaxation. We found that as the viscosity contrast increased, the time scale for shallow-shelf relaxation converged on the same time scale as for constant viscosity, except that the viscosity was replaced by the depth-averaged viscosity. This result follows from theory developed by \citet{Peg2012}. We also found that as the viscosity contrast increased, the onset of the shallow-shelf regime was delayed to larger scales of deformation, which yielded an empirical scaling between the viscosity contrast and the transition wavelength. These results show that for icy satellites with large viscosity contrasts (e.g. Enceladus), the deformation is primarily in a shallow-ice regime rather than shallow shelf, as previously assumed. Since the rate of topographic relaxation is important for inferences about gravity and ice thickness, these results place prior model assumptions and their conclusions on a firmer foundation for future studies of ice shell evolution at ocean worlds.   

\section{Acknowledgements}
Thanks to Sam Birch, Bill McKinnon, Francis Nimmo, and Georgia Peterson for insightful conversations at AGU. CRM, JJB, and TCT were supported by NASA (80NSSC21M0329, 80NSSC21K1804); CRM and AS acknowledge NSF (2012958); and CRM was also funded by ARO (78811EG). ACB was supported by NASA (80NSSC22K1318).


\bibliographystyle{plainnat}
\bibliography{Glib}


\end{document}